\documentclass{llncs}
\usepackage[utf8]{inputenc}
\usepackage{color}
\usepackage[setpagesize=false]{hyperref}
\usepackage{epsfig}

\usepackage{floatrow}
\newdimen\oldframetabcolsep
\newdimen\oldcolortabcolsep
\newdimen\oldpretabcolsep

\title{Reproducible and User-Controlled %
Software Environments in HPC with Guix}
\author{
Ludovic Court\`{e}s\inst{1}
 and Ricardo Wurmus\inst{2}
}
\institute{
Inria, Bordeaux, France 
\and
Max Delbrück Center for Molecular %
		    Medicine, Berlin, Germany 
}
\begin{document}
\date{}
\maketitle

\begin{abstract}
Support teams of high-performance computing (HPC) systems often %
find themselves between a rock and a hard place{\char58} on one hand, they %
understandably administrate these large systems in a conservative way, %
but on the other hand, they try to satisfy their users by deploying %
up-to-date tool chains as well as libraries and scientific software. %
HPC system users often have no guarantee that they will be able to %
reproduce results at a later point in time, even on the same %
system---software may have been upgraded, removed, or recompiled %
under their feet, and they have little hope of being able to %
reproduce the same software environment elsewhere.  We present %
GNU~Guix and the functional package management paradigm and show %
how it can improve reproducibility and sharing among researchers %
with representative use cases.
\end{abstract}

\section{Introduction}
\label{chapter14444}
HPC system administration has to satisfy two seemingly %
contradictory demands{\char58} on one hand administrators seek stability, which %
leads to a conservative approach to software management, and on the %
other hand users demand recent tool chains and huge scientific software %
stacks.  In addition, users often need different versions and different %
variants of a given software package.  To satisfy both, support teams %
end up playing the role of ``distribution maintainers''{\char58} they build and %
install tool chains, libraries, and scientific software packages %
manually---multiple variants thereof---and make them available %
{\textit{via}} ``environment modules''\cite{furlani1991:modules}, which %
allows users to pick the specific packages they want.\par
Unfortunately, software is often built and installed in an %
{\textit{ad hoc}} fashion, leaving users little hope of redeploying the %
same software environment on another system.  Worse, support teams %
occasionally have to remove installed software or to upgrade it in %
place, which means that users may eventually find themselves unable to %
reproduce their software environment, {\em{even on the same %
system}}.\par
Recently-developed tools such as EasyBuild \cite{geimer2014:easybuild} and Spack \cite{gamblin:spack-web} address part of the problem by automating package %
builds, supporting non-root users, and adding facilities to create %
package variants.  However, %
these tools fall short when it comes to build reproducibility.  First, %
build processes can trivially refer to tools or libraries already %
installed on the system.  Second, %
the {\textit{ad hoc}} naming conventions they rely on to identify builds %
fail to capture the directed acyclic graph (DAG) of dependencies that %
led to this particular build.\par
GNU~Guix is a general-purpose package manager that %
implements the functional package management paradigm pioneered by Nix %
\cite{courtes2013:functional,dolstra2004:nix}.  Many of its %
properties and features make it attractive in a multi-user HPC context{\char58} %
per-user profiles, transactional upgrades and roll-backs, and, more %
importantly, a controlled build environment to maximize reproducibility.\par
\hyperref[rationale]{Section~2} details our motivations. %
\hyperref[functional]{Section~3} describes the functional %
package management paradigm, its implementation in Guix, its impact on %
reproducibility, and how it can be applied to HPC systems.  \hyperref[use-cases]{Section~4} gives concrete use cases where Guix %
empowers users while guaranteeing reproducibility and sharing, while %
\hyperref[limitations]{Section~5} discusses limitations and %
remaining challenges.  Finally, \hyperref[related]{Section~6} compares to related work, and \hyperref[conclusion]{Section~7} concludes.\par

\section{Rationale}
\label{rationale}
Recent work on reproducible research insufficiently takes %
software environment reproducibility into account.  For example, the %
approach for verifiable computational results described in \cite{gavish2011:universal} focuses on workflows and conventions but does not %
mention the difficulty of reproducing full software environments. %
Likewise, the new Replicated Computational Results (RCR) initiative of %
the ACM Transactions on Mathematical Software acknowledges the %
importance of reproducible results, but does not adequately address the %
issue of software environments, which is a prerequisite. %
The authors of \cite{stanisic2014:effective} propose a %
methodology for reproducible research experiments in HPC.  To address %
the software-environment reproducibility problem they propose two %
unsatisfying approaches{\char58} one is to write down the %
version numbers of the dependencies being used, which is insufficient, %
and the other is to save and reuse full system images, which %
makes verifiability %
impractical---peers would have to download large images and would be %
unable to combine them with their own software environment.\par
Yet, common practices on HPC systems hinder reproducibility. %
For understandable stability reasons, %
HPC systems often run old GNU/Linux distributions that are rarely %
updated.  Thus, packages provided by the distribution are largely %
dismissed.  Instead support teams install packages from third-party %
repositories---but then they clobber the global {\texttt{/\-usr}} %
prefix, which sysadmins may want to keep under control, or install %
them from source by %
themselves and make them available through environment modules %
\cite{furlani1991:modules}.  Modules allow users to choose different %
versions or variants of the packages they use without interfering with %
each other.  However, when installed software is updated in place or %
removed, users suddenly find themselves unable to reproduce the software %
environment they were using.  Given these practices, reproducing the %
exact same software environment on a {\em{different}} HPC system %
seems out of reach.  It is nonetheless a very important property{\char58} It %
would allow users to assess the impact of the hardware on the software's %
performance---something that is very valuable in particular for %
developers of run-time systems---and it would allow other researchers to %
reproduce experiments on their system.\par
Essentially, by deploying software and environment modules, %
HPC support teams find themselves duplicating the work of GNU/Linux %
distributions, but why is that?  Historical package managers such as %
APT and RPM suffer from %
several limitations.  First, package binaries that every user installs, %
such as {\texttt{.deb}} files, are actually built on the package %
maintainer's machine, and details about the host %
may leak into the binary that is uploaded---a shortcoming that is now %
being addressed (see \hyperref[related]{Section~6}.)\par
Second, while it is in theory possible for a user to define %
their own variant of a package, as is often needed in HPC, this %
is often difficult in practice.  Users of RPM-based systems, %
for example, may be able to customize a {\texttt{.spec}} file to %
build a custom, relocatable RPM package, but only the administrator can %
install the package alongside its dependencies and register it in the %
central {\texttt{yumdb}} package database.  The %
lower-level {\texttt{rpm}} tool can use a separate package %
registry, which could be useful for unprivileged users; however RPM %
package databases cannot %
be composed, so users would need to manually track down and register the %
complete graph of dependencies, which is impractical at best.\par
Third, these tools implement an {\em{imperative}} and {\em{stateful}} package management model %
\cite{dolstra2004:nix}.  It is %
imperative in the sense that it modifies the set of available packages %
in place.  For example, switching to an %
alternative MPI implementation, or upgrading the OpenMP run-time library %
means that suddenly all the installed applications and libraries start %
using them.  It is stateful in the sense that the system state after a %
package management operation depends on its previous state.  Namely, the %
system state at a given point in time is the result of the series of %
installation and upgrade operations that have been made over time, and %
there may be no way to reproduce the exact same state elsewhere.  These %
properties are a serious hindrance to reproducibility.\par

\section{Functional Package Management}
\label{functional}
{\textit{Functional paradigm.}} Functional package management is %
a discipline that transcribes the functional programming paradigm to %
software deployment{\char58} build and installation processes are viewed as pure %
functions in the mathematical sense---whose result depends %
exclusively on the inputs---, and their result is a value---that %
is, an immutable directory.  Since build and installation processes are %
pure functions, their results can effectively be ``cached'' on %
disk.  Likewise, two %
independent runs of a given build process for a given set of inputs %
should return the same value---{\textit{i.e.}}, bit-identical %
files.  This approach was first described and implemented in the Nix %
package manager \cite{dolstra2004:nix}.  Guix reuses low-level %
mechanisms from Nix to implement the same paradigm, but offers a unified %
interface for package definitions and their implementations, all embedded %
in a single programming language \cite{courtes2013:functional}.\par
An obvious challenge is the implementation of this paradigm{\char58} %
How can build and install processes be viewed as pure?  To obtain that %
property, Nix and Guix ensure tight control over the build environment. %
In both cases, build processes are started by a privileged daemon, which %
always runs them in ``containers'' as implemented by the kernel Linux; %
that is, they run in a chroot environment, under a dedicated user ID, %
with a well-defined set of environment variables, %
with separate name spaces for PIDs, inter-process communication (IPC), %
networking, and so on.  The chroot environment contains %
only the directories corresponding to the explicitly declared inputs. %
This ensures that the build process cannot %
inadvertently end up using tools or libraries that it is not supposed to %
use.  The separate name spaces ensure that the build process cannot %
communicate with the outside world. %
Although it is not perfect as we will see in \hyperref[limitations]{Section~5}, this technique gives us confidence that build %
processes can indeed be viewed as pure functions, with  %
reproducible results.\par
Each build process produces one or more files in directories %
stored in a common place called {\em{the store}}, typically the {\texttt{/\-gnu/\-store}} directory.  Each entry in {\texttt{/\-gnu/\-store}} has a name %
that includes a hash of {\em{all the inputs}} of the build process %
that led to it.  By ``all %
the inputs'', we really mean all of them{\char58} This includes of course %
compilers and libraries, including the C library, but also build %
scripts and environment variable values.  This is recursive{\char58} The %
compiler's own directory name is a hash of the tools and libraries used %
to build, and so on, up to a set of pre-built binaries used %
for bootstrapping purposes---which can in turn be rebuilt using %
Guix \cite{courtes2013:functional}.  Thus, for each package that %
is built, the system has access to the {\em{complete DAG}} of %
dependencies used to build it.\par
\begin{figure}[ht]
\setlength{\oldpretabcolsep}{\tabcolsep}
\addtolength{\tabcolsep}{-\tabcolsep}
{\setbox1 \vbox \bgroup
{\noindent \texttt{\begin{tabular}{l}
{\textit{\ \ 1{\char58}\ }}({\textcolor[rgb]{0.4117647058823529,0.34901960784313724,0.8117647058823529}{{\textbf{define}}}}\ openmpi\\
{\textit{\ \ 2{\char58}\ }}\ \ (package\\
{\textit{\ \ 3{\char58}\ }}\ \ \ \ (name\ {\textcolor[rgb]{1.0,0.0,0.0}{"openmpi"}})\\
{\textit{\ \ 4{\char58}\ }}\ \ \ \ (version\ {\textcolor[rgb]{1.0,0.0,0.0}{"1.8.1"}})\\
{\textit{\ \ 5{\char58}\ }}\ \ \ \ (source\ (origin\\
{\textit{\ \ 6{\char58}\ }}\ \ \ \ \ \ \ \ \ \ \ \ \ \ (method\ url-fetch)\\
{\textit{\ \ 7{\char58}\ }}\ \ \ \ \ \ \ \ \ \ \ \ \ \ (uri\ (string-append\\
{\textit{\ \ 8{\char58}\ }}\ \ \ \ \ \ \ \ \ \ \ \ \ \ \ \ \ \ \ \ {\textcolor[rgb]{1.0,0.0,0.0}{"http{\char58}//www.open-mpi.org/software/ompi/v"}}\\
{\textit{\ \ 9{\char58}\ }}\ \ \ \ \ \ \ \ \ \ \ \ \ \ \ \ \ \ \ \ (version-major+minor\ version)\\
{\textit{\ 10{\char58}\ }}\ \ \ \ \ \ \ \ \ \ \ \ \ \ \ \ \ \ \ \ {\textcolor[rgb]{1.0,0.0,0.0}{"/downloads/openmpi-"}}\ version\ {\textcolor[rgb]{1.0,0.0,0.0}{".tar.bz2"}}))\\
{\textit{\ 11{\char58}\ }}\ \ \ \ \ \ \ \ \ \ \ \ \ \ (sha256\\
{\textit{\ 12{\char58}\ }}\ \ \ \ \ \ \ \ \ \ \ \ \ \ \ (base32\\
{\textit{\ 13{\char58}\ }}\ \ \ \ \ \ \ \ \ \ \ \ \ \ \ \ {\textcolor[rgb]{1.0,0.0,0.0}{"13z1q69f3qwmmhpglarfjminfy2yw4rfqr9jydjk5507q3mjf50p"}}))))\\
{\textit{\ 14{\char58}\ }}\ \ \ \ (build-system\ gnu-build-system)\ \ \ \ \ \ \ \ \ \ \ {\textcolor[rgb]{0.4666666666666667,0.4,0.0}{{\textbf{}}}}\\
{\textit{\ 15{\char58}\ }}\ \ \ \ (inputs\ `(({\textcolor[rgb]{1.0,0.0,0.0}{"hwloc"}}\ ,hwloc)\ \ \ \ \ \ \ \ \ \ \ \ \ \ \ \ {\textcolor[rgb]{0.4666666666666667,0.4,0.0}{{\textbf{}}}}\\
{\textit{\ 16{\char58}\ }}\ \ \ \ \ \ \ \ \ \ \ \ \ \ ({\textcolor[rgb]{1.0,0.0,0.0}{"gfortran"}}\ ,gfortran-4.8)\\
{\textit{\ 17{\char58}\ }}\ \ \ \ \ \ \ \ \ \ \ \ \ \ ({\textcolor[rgb]{1.0,0.0,0.0}{"pkg-config"}}\ ,pkg-config)))\\
{\textit{\ 18{\char58}\ }}\ \ \ \ (arguments\ '(\#{\char58}configure-flags\ `({\textcolor[rgb]{1.0,0.0,0.0}{"--enable-oshmem"}})))\\
{\textit{\ 19{\char58}\ }}\ \ \ \ (home-page\ {\textcolor[rgb]{1.0,0.0,0.0}{"http{\char58}//www.open-mpi.org"}})\\
{\textit{\ 20{\char58}\ }}\ \ \ \ (synopsis\ {\textcolor[rgb]{1.0,0.0,0.0}{"MPI-2\ implementation"}})\\
{\textit{\ 21{\char58}\ }}\ \ \ \ (description\ {\textcolor[rgb]{1.0,0.0,0.0}{"This\ is\ an\ MPI-2\ implementation\ etc."}})\\
{\textit{\ 22{\char58}\ }}\ \ \ \ (license\ bsd-2)))\\
\end{tabular}
}}
\egroup{\box1}}%
\setlength{\tabcolsep}{\oldpretabcolsep}
\caption{\label{fig-recipe}Guix package recipe of Open~MPI.}\end{figure}
Package recipes in Guix are written in a domain-specific %
language (DSL) embedded in the Scheme programming language.  \hyperref[fig-recipe]{Figure~1} shows, as an example, the recipe to %
build the Open~MPI library.  The {\texttt{package}} form evaluates to a {\em{package %
object}}, which is just a ``regular'' Scheme value; the {\texttt{define}} %
form defines the {\texttt{openmpi}} variable to hold that value.\par
\begin{figure}[ht]
\setlength{\oldpretabcolsep}{\tabcolsep}
\addtolength{\tabcolsep}{-\tabcolsep}
{\setbox1 \vbox \bgroup
{\noindent \texttt{\begin{tabular}{l}
{\textcolor[rgb]{0.4666666666666667,0.4,0.0}{{\textbf{;;\ Query\ the\ direct\ and\ indirect\ inputs\ of\ Open\ MPI.}}}}\\
{\textcolor[rgb]{0.4666666666666667,0.4,0.0}{{\textbf{;;\ Each\ input\ is\ represented\ by\ a\ label/package\ tuple.}}}}\\
(map\ (match-lambda\\
\ \ \ \ \ \ \ ((label\ package)\\
\ \ \ \ \ \ \ \ (package-full-name\ package)))\\
\ \ \ \ \ (package-transitive-inputs\ openmpi))\\
\end{tabular}
}}
\egroup{\box1}}%
\setlength{\tabcolsep}{\oldpretabcolsep}
... yields{\char58}\par
\setlength{\oldpretabcolsep}{\tabcolsep}
\addtolength{\tabcolsep}{-\tabcolsep}
{\setbox1 \vbox \bgroup
{\noindent \texttt{\begin{tabular}{l}
({\textcolor[rgb]{1.0,0.0,0.0}{"hwloc-1.10.1"}}\ {\textcolor[rgb]{1.0,0.0,0.0}{"gfortran-4.8.5"}}\ {\textcolor[rgb]{1.0,0.0,0.0}{"pkg-config-0.28"}})\\
\end{tabular}
}}
\egroup{\box1}}%
\setlength{\tabcolsep}{\oldpretabcolsep}
\caption{\label{fig-query}Querying the dependencies of a package object.}\end{figure}
Line 14 specifies that the package is to be built %
according to the GNU standards---{\textit{i.e.}}, %
the well-known {\texttt{.\-/\-configure \&\& make \&\& make install}} sequence %
(similarly, Guix defines {\texttt{cmake-build-system}}, and so on.) %
The {\texttt{inputs}} field on line 15 specifies %
the direct dependencies of the package.  The field refers to the %
{\texttt{hwloc}}, {\texttt{gfortran-4.\-8}}, and {\texttt{pkg-config}} variables, %
which are also bound to package objects (their definition is not shown %
here.)  It would be inconvenient to specify all the standard inputs, %
such as Make, GCC, Binutils so these are implicit here; %
as it compiles package objects to a lower-level intermediate %
representation, {\texttt{gnu-build-system}} automatically inserts %
references to specific package objects for GCC, Binutils, etc. %
Since we are manipulating ``normal'' Scheme objects, we can use the API %
of Guix to query those package objects, as illustrated with the code in %
\hyperref[fig-query]{Figure~2}, which queries the name and %
version of the direct and indirect dependencies of our package\footnote{This document is an ``active paper'' written in Skribilo, a Scheme-based authoring %
tool, which allows us to use Guix and run this code from the document.}.\par
With that definition in place, running {\texttt{guix build %
openmpi}} returns the directory name %
{\texttt{/\-gnu/\-store/\-rmnib3ggm0dq32ls160ja882vanb69fi-openmpi-1.\-8.\-1}}.  If %
that directory did not already exist, the daemon spawns the build %
process in its isolated environment with write access to this directory. %
Of course users never have to type these long {\texttt{/\-gnu/\-store}} file %
names.  They can install packages in their {\em{profile}} using %
the {\texttt{guix package}} command, which essentially creates symbolic %
links to the selected {\texttt{/\-gnu/\-store}} items.   By default, the tree of %
symbolic links is rooted at {\texttt{$_{\mbox{\char126}}$/\-.\-guix-profile}}, but users %
can also create independent profiles in %
arbitrary places of the file system.  For instance, a user may choose to %
have GCC and Open~MPI in the default profile, and to populate %
another profile with Clang and MPICH2.\par
It is then a matter of defining %
the search paths for the compiler, linker, and other tools {\textit{via}} %
environment variables.  Fortunately, Guix keeps track of that and the %
{\texttt{guix package --search-paths}} command returns all the necessary %
environment variable definitions in Bourne shell syntax.  For example, %
when both the GCC tool chain and Open~MPI are installed, the command %
returns definitions for the {\texttt{PATH}}, {\texttt{CPATH}}, and {\texttt{LIBRARY\_PATH}} environment variables, and these definitions can be %
passed to the {\texttt{eval}} shell built-in command.\par

\section{Use Cases}
\label{use-cases}
We explore practical use cases where Guix improves %
experimentation reproducibility for a user of a given system, supports %
the deployment of complex software stacks, allows a software environment %
to be replicated on another system, and finally allows fine %
customization of the software environment.\par

\subsection{Usage Patterns on an HPC Cluster}
\label{section14811}
One of the key features of Guix and Nix is that they %
securely permit unprivileged users to install packages in the store %
\cite{dolstra2004:nix}.  To build a package, the {\texttt{guix}} %
commands connect to the build daemon, which then performs the build (if %
needed) on their behalf, in the isolated environment. %
When two users build the exact same package, both end up using the exact %
same {\texttt{/\-gnu/\-store}} file name, and storage is shared.  If a user %
tries to build, say, a malicious version of the C library, then the %
other users on the system will not use it, simply because they cannot %
guess its {\texttt{/\-gnu/\-store}} file name---unless %
they themselves explicitly build the very same modified C library.\par
Guix is deployed at the Max Delbrück Center for Molecular Medicine (MDC), %
Berlin, where the store is shared among 250 cluster %
nodes and an increasing number of user workstations.  It is now gradually %
replacing other methods %
of software distribution, such as statically linked binaries on group %
network shares, relocatable RPMs installed into group prefixes, %
one-off builds on the cluster, and user-built software %
installed in home directories. %
The researchers use tens of bioinformatics tools as well as %
frameworks such as Biopython, NumPy, SciPy, and SymPy. %
The functional packaging approach proved particularly %
useful in the ongoing efforts to move dozens of users and their custom software %
environments from an older cluster running Ubuntu to a new cluster %
running a version of CentOS, because software packaged with Guix does %
not depend on any of the host system's libraries and thus can be used %
on very different systems without any changes to the packages. %
Research groups now have a shared profile for common %
applications, whereas individual users can manage their own %
profiles for custom software, legacy versions of bioinformatics tools %
to reproduce published results, bleeding-edge tool chains, or even for %
complete workflows.\par
\begin{figure}[ht]
\setlength{\oldpretabcolsep}{\tabcolsep}
\addtolength{\tabcolsep}{-\tabcolsep}
{\setbox1 \vbox \bgroup
{\noindent \texttt{\begin{tabular}{l}
{\textcolor[rgb]{0.4666666666666667,0.4,0.0}{{\textbf{;;\ This\ file\ can\ be\ passed\ to\ 'guix\ package\ --manifest'.}}}}\\
(use-modules\ (gnu\ packages\ base)\ (gnu\ packages\ gcc)\\
\ \ \ \ \ \ \ \ \ \ \ \ \ (my-openmpi))\\
\\
(packages-$>$manifest\\
\ (list\ glibc-utf8-locales\ gnu-make\ gcc-toolchain\ openmpi))\\
\end{tabular}
}}
\egroup{\box1}}%
\setlength{\tabcolsep}{\oldpretabcolsep}
\caption{\label{fig-manifest}Declaring the set of packages to be installed in a %
            profile.}\end{figure}
Guix supports two ways to manage a %
profile.  The first one is to make transactions that add, upgrade, or %
remove packages in the profile{\char58} {\texttt{guix package --install openmpi %
--remove mpich2}} installs Open~MPI and removes MPICH2 %
in a single transaction that can be rolled back.  The %
second approach is to {\em{declare}} the desired contents of the %
profile and make that effective{\char58} the user writes in a file a code %
snippet that lists the requested packages (see \hyperref[fig-manifest]{Figure~3}) and then runs {\texttt{guix package %
--manifest=my-packages.\-scm}}.\par
This declarative profile management makes it easy to %
replicate a profile, but it is symbolic{\char58} It uses whatever package %
objects the variables are bound to ({\texttt{gnu-make}}, {\texttt{gcc-toolchain}}, etc.), but these variables are typically defined in %
the {\texttt{(gnu packages …)}} modules that Guix comes with.  Thus the %
precise packages being installed depend on the version of Guix that is %
available.  Specifying the Git commit of Guix %
in addition to the declaration in %
\hyperref[fig-manifest]{Figure~3} is all it takes to %
reproduce the exact same {\texttt{/\-gnu/\-store}} items.\par
Another approach to achieve bit-identical reproduction of a %
user's profile is by saving the contents of its transitive closure using {\texttt{guix %
archive --export}}.  The resulting archive can be transferred to another %
system and restored at any point in time using {\texttt{guix archive %
--import}}.  This should significantly facilitate experimentation and %
sharing among peers.\par

\subsection{Customizing Packages}
\label{customizing}
Our colleagues at Inria in the HiePACS and Runtime teams %
develop a complete linear algebra software stack going from sparse %
solvers such as PaStiX and dense solvers such as Chameleon, %
to run-time support libraries and compiler extensions such as %
StarPU\footnote{\href{http://starpu.gforge.inria.fr/}{{\texttt{http{\char58}/\-/\-starpu.\-gforge.\-inria.\-fr/\-}}}} and hwloc.  While %
developers of simulations want to be able to deploy the whole stack, %
developers of solvers only need their project's %
dependencies, possibly several variants thereof.  For instance, %
developers of Chameleon may want to test their software against several %
versions of StarPU, or against variants of StarPU built with different %
compile-time options.  Finally, developers of the lower-level layers, %
such as StarPU, may want to test the effect of changes they make on %
higher-level layers.\par
\begin{figure}[ht]
\setlength{\oldpretabcolsep}{\tabcolsep}
\addtolength{\tabcolsep}{-\tabcolsep}
{\setbox1 \vbox \bgroup
{\noindent \texttt{\begin{tabular}{l}
({\textcolor[rgb]{0.4117647058823529,0.34901960784313724,0.8117647058823529}{{\textbf{define}}}}\ starpu-1.2rc\ \ \ \ \ \ \ \ \ \ \ \ \ \ {\textcolor[rgb]{0.4666666666666667,0.4,0.0}{{\textbf{;release\ candidate}}}}\\
\ \ (package\ (inherit\ starpu)\\
\ \ \ \ (version\ {\textcolor[rgb]{1.0,0.0,0.0}{"1.2.0rc2"}})\\
\ \ \ \ (source\ (origin\\
\ \ \ \ \ \ \ \ \ \ \ \ \ (method\ url-fetch)\\
\ \ \ \ \ \ \ \ \ \ \ \ \ (uri\ (string-append\ {\textcolor[rgb]{1.0,0.0,0.0}{"http{\char58}//starpu.gforge.inria.fr/files/"}}\\
\ \ \ \ \ \ \ \ \ \ \ \ \ \ \ \ \ \ \ \ \ \ \ \ \ \ \ \ \ \ \ \ \ {\textcolor[rgb]{1.0,0.0,0.0}{"starpu-"}}\ version\ {\textcolor[rgb]{1.0,0.0,0.0}{".tar.gz"}}))\\
\ \ \ \ \ \ \ \ \ \ \ \ \ (sha256\\
\ \ \ \ \ \ \ \ \ \ \ \ \ \ (base32\\
\ \ \ \ \ \ \ \ \ \ \ \ \ \ \ {\textcolor[rgb]{1.0,0.0,0.0}{"0qgb6yrh3k745grjj14gc2vl6a99m0ljcsisfzcwyhg89vdpx42v"}}))))))\\
\\
({\textcolor[rgb]{0.4117647058823529,0.34901960784313724,0.8117647058823529}{{\textbf{define}}}}\ starpu-with-simgrid\\
\ \ (package\ (inherit\ starpu)\\
\ \ \ \ (name\ {\textcolor[rgb]{1.0,0.0,0.0}{"starpu-with-simgrid"}})\ \ {\textcolor[rgb]{0.4666666666666667,0.4,0.0}{{\textbf{;name\ shown\ in\ the\ user\ interface}}}}\\
\ \ \ \ (inputs\ `(({\textcolor[rgb]{1.0,0.0,0.0}{"simgrid"}}\ ,simgrid)\\
\ \ \ \ \ \ \ \ \ \ \ \ \ \ ,@(package-inputs\ starpu)))))\\
\end{tabular}
}}
\egroup{\box1}}%
\setlength{\tabcolsep}{\oldpretabcolsep}
\caption{\label{fig-variants}Defining variants of the default recipe for %
            StarPU.}\end{figure}
This use case leads to two requirements{\char58} that users be able %
to customize and non-ambiguously specify a package DAG, and that they be %
able to reproduce any variant of their package DAG.  Guix allows them to %
define variants; the code for these variants can be stored in a %
repository of their own and made visible to the {\texttt{guix}} %
commands by defining the {\texttt{GUIX\_PACKAGE\_PATH}} environment variable. %
\hyperref[fig-variants]{Figure~4} shows an example of such %
package variants{\char58} based on the pre-existing {\texttt{starpu}} variable, the %
first variant defines a package for a new StarPU release candidate, %
simply by changing its {\texttt{source}} field, while the second variant %
adds the optional dependency on the SimGrid simulator---a variant %
useful to scheduling practitioners, but not necessarily to solver %
developers.\par
These StarPU package definitions are obviously useful to %
users of StarPU{\char58} They can install them with {\texttt{guix package -i %
starpu}} and similar commands.  But they are also useful to StarPU %
developers{\char58} They can enter a ``pristine'' development environment %
corresponding to the dependencies given in the recipe by running {\texttt{guix environment starpu --pure}}.  This command spawns a shell where %
the usual {\texttt{PATH}}, {\texttt{CPATH}} etc. environment variables are %
redefined to refer precisely to the inputs specified in the recipe.  This %
amounts to creating a profile on the fly, containing only the tools and %
libraries necessary when developing StarPU.  This is notably useful when %
dealing with build systems that support optional dependencies.\par
\begin{figure}[ht]
\setlength{\oldpretabcolsep}{\tabcolsep}
\addtolength{\tabcolsep}{-\tabcolsep}
{\setbox1 \vbox \bgroup
{\noindent \texttt{\begin{tabular}{l}
\ \ \ ({\textcolor[rgb]{0.4117647058823529,0.34901960784313724,0.8117647058823529}{{\textbf{define}}}}\ (make-chameleon\ name\ starpu)\\
\ \ \ \ \ (package\\
\ \ \ \ \ \ \ (name\ name)\\
\ \ \ \ \ \ \ {\textcolor[rgb]{0.4666666666666667,0.4,0.0}{{\textbf{;;\ {\char91}other\ fields\ omitted{\char93}}}}}\\
\ \ \ \ \ \ \ (inputs\ `(({\textcolor[rgb]{1.0,0.0,0.0}{"starpu"}}\ ,starpu)\\
\ \ \ \ \ \ \ \ \ \ \ \ \ \ \ \ \ ({\textcolor[rgb]{1.0,0.0,0.0}{"blas"}}\ ,atlas)\ ({\textcolor[rgb]{1.0,0.0,0.0}{"lapack"}}\ ,lapack)\\
\ \ \ \ \ \ \ \ \ \ \ \ \ \ \ \ \ ({\textcolor[rgb]{1.0,0.0,0.0}{"gfortran"}}\ ,gfortran-4.8)\\
\ \ \ \ \ \ \ \ \ \ \ \ \ \ \ \ \ ({\textcolor[rgb]{1.0,0.0,0.0}{"python"}}\ ,python-2)))))\\
\\
\ \ \ ({\textcolor[rgb]{0.4117647058823529,0.34901960784313724,0.8117647058823529}{{\textbf{define}}}}\ chameleon\\
\ \ \ \ \ (make-chameleon\ {\textcolor[rgb]{1.0,0.0,0.0}{"chameleon"}}\ starpu))\\
\ \ \ ({\textcolor[rgb]{0.4117647058823529,0.34901960784313724,0.8117647058823529}{{\textbf{define}}}}\ chameleon/starpu-simgrid\\
\ \ \ \ \ (make-chameleon\ {\textcolor[rgb]{1.0,0.0,0.0}{"chameleon-simgrid"}}\ starpu-with-simgrid))\\
\end{tabular}
}}
\egroup{\box1}}%
\setlength{\tabcolsep}{\oldpretabcolsep}
\caption{\label{fig-chameleon}Defining a function that returns a package object %
for the Chameleon solver.}\end{figure}
Now that we have several StarPU variants, we %
want to allow direct %
and indirect users to select the variant that they want.  A simple way to %
do that is to write, say, a function that takes a %
{\texttt{starpu}} parameter and returns a package that uses it as its input %
as show in \hyperref[fig-chameleon]{Figure~5}.  To allow users to refer %
to one or the other variant at the command line, we use different %
values for the {\texttt{name}} field.\par
This approach is reasonable when there is a small number of %
variants, but it does not scale to more complex DAGs.  As an example, %
StarPU can be built with MPI support, in which case Chameleon also needs %
to be explicitly linked against the same MPI implementation. One way to do that %
is by writing a function that recursively adjusts the package labeled %
{\texttt{"mpi"}} in the {\texttt{inputs}} field of packages in the DAG. %
 No matter how complex the transformations are, a package %
object unambiguously represents a reproducible build process.  In that %
sense, Guix allows environments to be reproduced at different sites, or %
by different users, while still supporting users needing complex %
customization.\par

\section{Limitations and Challenges}
\label{limitations}
{\em{Privileged daemon.}} Nix and Guix address many of the %
reproducibility issues encountered in package deployment, and Guix %
provides APIs that can facilitate the %
development of package variants as is useful in HPC.  Yet, to our %
knowledge, neither Guix nor Nix are widely deployed on HPC systems.  An %
obvious reason that limits adoption is the requirement to have the build %
daemon run with root privileges---without which it would be unable %
to use the Linux kernel container facilities that allow it to isolate %
build processes and maximize build reproducibility.  System %
administrators are wary of installing privileged daemons, and so HPC %
system users trade reproducibility for practical approaches.\par
{\em{Cluster setup.}} All the {\texttt{guix}} commands are %
actually clients of the daemon.  In a typical cluster setup, system %
administrators may want to run a single daemon on one specific node and %
to share {\texttt{/\-gnu/\-store}} among all the nodes.  At the time of %
writing, Guix does not yet allow communication with a remote daemon. %
For this reason, Guix users at the MDC are required to manage their %
profiles from a specific node; other nodes can use the profiles, but not %
modify them.  Allowing the {\texttt{guix}} commands to communicate with a %
remote daemon will address this issue.\par
Additionally, compute nodes typically lack %
access to the Internet.  However, the daemon needs to be able to download %
source code tarballs or pre-built binaries from external servers.  Thus, %
the daemon must run on a node with Internet access, which could be %
contrary to the policy on some clusters.\par
{\em{OS kernel.}} By choosing not to use a full-blown VM %
and thus relying on the host OS kernel, our system assumes that the %
kernel interface is stable and that the kernel has little or no impact %
on program behavior.  While this may sound like a broad assumption, our %
experience has shown that it holds for almost all the software packages %
provided by Guix.  Indeed, while applications may be sensitive to %
changes in the C library, only low-level kernel-specific user-land %
software is really sensitive to changes in the kernel.  The build daemon %
itself relies on features that have been available in the kernel for %
several years.\par
{\em{Non-determinism.}} Despite the use of %
isolated containers to run build processes, there are still a few sources %
of non-determinism that build systems of packages might use and %
that can impede reproducibility.  In particular, %
details about the operating system kernel and the hardware being used %
can ``leak'' to build processes.  For example, the kernel Linux provides %
system calls such as {\texttt{uname}} and interfaces such as %
{\texttt{/\-proc/\-cpuinfo}} that leak information about the host; independent %
builds on different hosts could lead to different results if this %
information is used.  Likewise, the {\texttt{cpuid}} instruction leaks %
hardware details.\par
Fortunately, few software packages depend on this information. %
Yet, the proportion of packages depending on it is higher in the HPC %
world.  A notable example is the ATLAS linear algebra system, which %
fine-tunes itself based on details about the CPU micro-architecture. %
Similarly, profile-guided optimization (PGO), where the compiler %
optimizes code based on a profile gathered in a previous run, undermines %
reproducibility.  Running build processes in full-blown VMs %
would address some of these issues, but with a potentially %
significant impact on build performance, and possibly preventing %
important optimization techniques in the HPC context.\par
{\em{Proprietary software.}} GNU~Guix does not provide %
proprietary software packages.  Unfortunately, proprietary software is %
still relatively common in HPC, be it linear algebra libraries or GPU %
support.  Yet, we see it as a strength more than a limitation.  Often, %
these ``black boxes'' inherently limit reproducibility---how is one %
going to reproduce a software environment without permission %
to run the software in the first place?  What if the software %
depends on the ability to ``call home'' to function at all?  More %
importantly, we view reproducible software environments and reproducible %
science as a tool towards improved and shared knowledge; %
developers who deny the freedom to study and modify their code work %
against this goal.\par

\section{Related Work}
\label{related}
{\em{Reproducible builds.}} Reproducible software %
environments have only recently become an active research area. %
One of the earliest pieces of work in this area is the Vesta software configuration %
system \cite{heydon2000:caching}.  Vesta provides a DSL that %
allows users to describe build operations, similar to Nix %
\cite{dolstra2004:nix}. More recently, projects such as Debian's Reproducible, Fedora's %
Mock, or Gitian have intended to improve reproducibility and %
verifiability of mainstream package distributions. %
Google's recent Bazel build tool relies on container facilities provided %
by the kernel Linux and provides another DSL to describe build %
operations.\par
Reproducibility can be achieved with heavyweight approaches %
such as full operating system deployments, be it on hardware or in VMs %
or containers \cite{boettiger2015:docker,vangorp2011:share,jeanvoine2013:kadeploy3,ruiz2015:kameleon}.  In addition to being %
resource-hungry, these approaches are coarse-grain and do not compose{\char58} %
if two different VM/container images or ``software appliances'' provide %
useful features or packages, the user has to make a binary choice and %
cannot combine the features or packages they offer.  Furthermore, %
``Docker files'', ``Vagrant files'', and Kameleon ``recipes'' \cite{ruiz2015:kameleon} suffer from being too broad for the purposes of %
reproducing a software environment---they are about configuring %
complete operating systems---and from offering an inappropriate level %
of abstraction---these recipes list commands to {\em{modify}} the %
state of the system image to obtain the desired state, whereas Guix %
allows users to {\em{declare}} the desired environment in terms of %
software packages.  Lastly, the tendency to rely on complete third-party %
system images is a security concern\footnote{``Over 30\% of Official %
Images in Docker Hub Contain High Priority Security Vulnerabilities'', %
\href{http://www.banyanops.com/blog/analyzing-docker-hub/}{{\texttt{http{\char58}/\-/\-www.\-banyanops.\-com/\-blog/\-analyzing-docker-hub/\-}}}.} %
Building upon third-party binary images also puts a barrier on %
reproducibility{\char58} Users may have recipes to rebuild their own software %
from source, but the rest of the system is essentially considered as a %
``black box'', which, if it can be rebuilt from source at all, can only %
be rebuilt using a completely different tool set.\par
{\em{HPC package management.}} In the HPC community, %
efforts have focused primarily on the automation of software deployment %
and the ability for users to customize their build environment %
independently of each other.  The latter has been achieved by %
``environment modules'', a simple but efficient tool set that is still %
widely used today \cite{furlani1991:modules}.  Build and %
deployment automation is more recent with the development of specialized %
package management tools such as EasyBuild \cite{geimer2014:easybuild} and Spack \cite{gamblin:spack-web}.\par
Both EasyBuild and Spack have the advantage of being %
installable by unprivileged users since they do not rely on privileged %
components, unlike Guix and Nix.  The downside is that they cannot use %
the kernel's container facilities, which seriously hinders %
build reproducibility.  When used in the user's home directories,  %
each user may end up %
rebuilding the same compiler, libraries, etc., which can be costly in %
terms of CPU, bandwidth, and disk usage.  Conversely, Nix and Guix %
support safe sharing of builds.\par
EasyBuild aims to support multiple package variants, such as %
packages built with different compilers, or linked against different MPI %
implementations.  To achieve that, it relies on directory naming %
conventions; for instance, {\texttt{OpenMPI/\-1.\-7.\-3-GCC-4.\-8.\-2}} contains %
packages built with the specified MPI implementation and compiler.  Such %
conventions fail to capture the full complexity of the DAG and %
configuration space.  For instance, the convention arbitrarily omits the %
C library, linker, or configuration flags being used.\par
EasyBuild is tightly integrated with environment modules \cite{furlani1991:modules}, which are familiar to most users of HPC %
systems.  While modules provide users with flexible environments, they %
implement an imperative, stateful paradigm{\char58} Users run a sequence of {\texttt{module load}} and {\texttt{module unload}} commands that {\em{alter}} %
the current environment.  This can make it much harder to reason about %
and reproduce an environment, as opposed to the declarative approaches %
implemented by {\texttt{guix package --manifest}} and {\texttt{guix %
environment}}.\par
Like EasyBuild and similarly to Guix, Spack implements build %
recipes as first-class objects in a general-purpose language, Python, %
which facilitates customization and the creation of package variants. %
In addition, Spack provides a rich command-line interface that allows %
users to express variants similar to those discussed in \hyperref[customizing]{Section~4.2}.  This appears to be very convenient for %
common cases, although there are limits to the expressivity and %
readability of such a compact syntax.\par

\section{Conclusion}
\label{conclusion}
Functional package managers provide the foundations for %
reproducible software environments, while still allowing fine-grain %
software composition and not imposing high disk and RAM costs. %
Today, GNU~Guix comes with %
2,060 packages, including many of the common %
HPC tools and libraries as well as around 50 bioinformatics packages. %
It is deployed on the clusters of the MDC Berlin,  %
and being discussed as one of the packaging options by %
the Open Bioinformatics Foundation, a non-profit for the biological %
research community.  We hope to see more HPC deployments of Guix in the %
foreseeable future.\par
GNU~Guix benefits from contributions by about 20 people %
each month.  It is the foundation of the Guix System Distribution, a %
standalone, reproducible GNU/Linux distribution.\par

\section*{Acknowledgments}
We would like to thank Florent Pruvost, Emmanuel Agullo, and %
Andreas Enge at Inria and Eric Bavier at Cray Inc. for insightful %
discussions and comments on an earlier draft.  We are grateful to the %
Guix contributors who keep improving the system.\par
{\small{\begin{flushleft}
\end{flushleft}
}}
\end{document}